\newcommand{\etal}{\emph{et al}.}

\newcommand{\minitab}[2][1]{\begin{tabular}{#1}#2\end{tabular}}
\documentclass[10pt, conference, compsocconf]{IEEEtran}
\usepackage{multirow}
\usepackage{pdflscape}
\usepackage{rotating}
\begin{document}
\date{}
\title{Towards a better understanding of testing if conditionals}

%\numberofauthors{4} %  in this sample file, there are a *total*
%% of EIGHT authors. SIX appear on the 'first-page' (for formatting
%% reasons) and the remaining two appear in the \additionalauthors section.
%%

\author{\IEEEauthorblockN{Shimul Kumar Nath\IEEEauthorrefmark{1},
Robert Merkel\IEEEauthorrefmark{2},
Man Fai Lau\IEEEauthorrefmark{1} and 
Tanay Kanti Paul\IEEEauthorrefmark{1} \\
\IEEEauthorblockA{\IEEEauthorrefmark{1}Faculty of Information and Communication Technologies, 
Swinburne University of Technology, Hawthorn, Australia \\
%Hawthorn, VIC. 3122, Australia \\
Email: \texttt{\{snath,elau,tpaul\}@swin.edu.au}}
\IEEEauthorblockA{\IEEEauthorrefmark{2}School of Information Technology, Monash University, Clayton, Australia \\
Email: \texttt{robert.merkel@monash.edu}}
\vspace{-17pt}
}
}

\maketitle
\begin{abstract}
Fault based testing is a technique in which test cases are chosen to reveal certain classes of faults.  At present, testing professionals use their personal experience to select testing methods for fault classes considered the most likely to be present.  However, there is little empirical evidence available in the open literature to support these intuitions.  By examining the source code changes when faults were fixed in seven open source software artifacts, we have classified bug fix patterns into fault classes, and recorded the relative frequencies of the identified fault classes.  This paper reports our findings related to ``if-conditional'' fixes.  We have classified the ``if-conditional'' fixes into fourteen fault classes and calculated their frequencies.  We found the most common fault class related to changes within a single ``atom''.  The next most common fault was the omission of an ``atom''.  We analysed these results in the context of Boolean specification testing.
\end{abstract}


%% A category with the (minimum) three required fields
%%A category including the fourth, optional field follows...
%\category{D.2.8}{Software Engineering}{Metrics}[complexity measures, performance measures]

%\keywords{ACM proceedings, \LaTeX, text tagging} % NOT required for Proceedings
%
\section{Introduction} \label{Introdution}

Unit testing remains the most common method to verify the quality and reliability of the modules and subsystems which make up a complex 
software system.  As such, the selection of the most effective unit testing techniques for the 
particular software under test is of much practical importance.  There are many testing techniques that can be used to select test cases for unit testing, therefore, testers need to use their judgement to choose which technique or techniques they will use to 
select test cases.  

One technique that software testers may apply is fault based 
testing.  Fault based testing~\cite{Larry:Fault} is to choose test cases that can uncover 
specific fault types, if they exist, in the software under test.  If the software passes 
the test cases, the testers can have confidence that those particular fault types do not exist in the 
software system.  Of course, no testing method other than exhaustive testing can guarantee the absence 
of \emph{all} faults; fault-based testing merely provides confidence that a specific class of faults is
not present.

For fault based testing to be useful, testers must select a fault-based technique that reveals 
faults which are actually present in the software system.  
As the faults that actually present in the software are unknown before 
testing, testers must predict which fault types are likely to be 
present.  The best way to inform such predictions is to collect empirical 
data about fault frequencies in real life software artifacts.

Recently Pan \etal~\cite{Pan:BugPattern} performed an empirical study of ``bug fix patterns'' - that is, patterns of source code changes seen 
when a bug is fixed - in seven open source Java artifacts.  They analysed the differences between two incremental updated versions stored on the repositories of these individual projects.  Pan \etal~categorized the identified changes into various bug fix patterns.  Their results show that there are two particularly common bug fix patterns: one is related to changes in method call parameters (e.g. \verb~methodName(ownerName, amount)~ changed to \verb~methodName(accountId, amount)~) and the other is related to changes in the conditional expressions in programming constructs such as those in  \verb+if+ conditionals (e.g. \verb~"if (x > 4)"~ changed to \verb~"if (x > 4 && y < 3)"~).
The method call parameter changes account for 14.9--25.5\% of the defined bug fix patterns in the seven systems studied by them whereas the changes in the \verb+if+ conditional expressions are 5.6--18.6\%.
Given the high frequency of bugs related to \verb+if+ conditional expressions, and given that there are some existing fault based testing techniques for testing boolean specification (e.g.~\cite{Chilenski:MCDC,YuLauChen:j-jss-06-1}), it is worthwhile investigating how often a particular types of these Boolean expressions related faults are made by the software developers. 

Pan \etal\ classified the \verb+if+ condition change (IF-CC) fix patterns into six categories.  However this classification is somewhat ad-hoc, not orthogonal (i.e, some of the fix patterns matched more than one category) and does not align with any ``pre-defined'' fault classes (e.g.~\cite{LauYu:j-tosem-05}).  As a result, testers may not be able to use this information to choose appropriate fault specific testing techniques.  
In this study, we have devised a classification scheme of the \verb+if+ condition change patterns and calculated their relative frequencies, which, as well as being orthogonal, aligns as closely as possible to the fault classes of Boolean expression testing methods~\cite{LauYu:j-tosem-05}.  

Our study was designed to provide testers additional information for choosing fault specific testing for \verb+if+ conditional faults.  In this context, we aimed to answer the following research questions:

%\begin{list}{$\bullet$}{\setlength{\leftmargin}{0.05in}}
%\item{What percentage of IF-CC fix patterns can be classified automatically based on the fault classes of Boolean expressions?}
%\item{What are the relative frequencies of the IF-CC fix patterns over several projects?}
%\end{list}

\begin{itemize}%{\setlength{\leftmargin}{0.05in}}
\item {\bf RQ1: }What percentage of IF-CC fix patterns can be classified automatically based on the fault classes of Boolean expressions?
\item {\bf RQ2: }What are the relative frequencies of the IF-CC fix patterns over several projects?
\end{itemize}

We believe that our results will enable software testers to make more informed decision when choosing testing methods.  In particular, we wished to 
examine the effectiveness of existing Boolean expression testing techniques in revealing the major fault classes relating to the IF-CC fix patterns.

The paper is organised as follows.  Section~\ref{Preliminaries} describes some important definitions used in this paper.  Section~\ref{Classification:if} describes the IF-CC fix patterns, LRF sub-patterns and non-fix patterns.  Section~\ref{Research:Method} presents the methodology that we applied to our research.  Section~\ref{tools:verification} describes the tools implementation and verification.  Section~\ref{Result} presents the extracted hunk data and frequencies of IF-CC fault classes.  Section~\ref{Discussion} examines issues relating to cross project similarity as well as the effectiveness of specification based testing of \verb+if+ conditional.  Section~\ref{Threat:validity} discusses threats to validity.  Section~\ref{Related:work} summarizes some related works.  Section~\ref{Conclusion:Future} reports our conclusions and suggestions for future works.

%\small
%table for if-cc fault class definition ---------------------------
\begin{table*}[!t]
\centering
\caption{IF-CC fix patterns and their corresponding examples }
%\vspace{-3pt}
%\begin{tabular}[c]{|p{4cm}|p{6cm}|l|l|}
\begin{tabular}[c]{|p{7cm}|p{10cm}|}
\hline
\textbf{IF-CC patterns } & \textbf{Example} \\
\hline
Literal Reference Fault (LRF): A single atom change in an if condition
	& \begin{minipage}[t]{10cm}
		\begin{verbatim}
		-if(data.getUsername() == null)
		+if(getUsername() == null)
		\end{verbatim}
        \end{minipage} \\
\hline

Literal Omission Fault (LOF):Addition of an atom with a leading logical AND (\&\&) in fix hunk
	& \begin{minipage}[t]{10cm}
		\begin{verbatim}
		-if(source == caretStatus)
		+if(source == caretStatus &&  evt.getClickCount() == 2)
		\end{verbatim}
	\end{minipage} \\ 
\hline

Term Omission Fault (TOF): Addition of an atom with a leading logical OR ($||$) in fix hunk
	& \begin{minipage}[t]{10cm}
		\begin{verbatim}
		-if(idealIndent == -1 || 
		-(!canDecreaseIndent && idealIndent < currentIndent))
		+if(idealIndent == -1 || 
		+idealIndent == currentIndent  || (!canDecreaseIndent	&&
		+idealIndent < currentIndent)) 
		\end{verbatim}
	\end{minipage} \\
\hline

Literal Insertion Fault (LIF): Removal of an atom in fix hunk that was in bug hunk with a leading logical AND ($\&\&$)
	& \begin{minipage}[t]{10cm}
		\begin{verbatim}
		-if(sorter != null && buffers.contains(buffer))
		+if(sorter != null)
		\end{verbatim}
	\end{minipage} \\
\hline

Term Insertion Fault (TIF): Removal of an atom in fix hunk that was in bug hunk with a leading logical OR ($||$)
	& \begin{minipage}[t]{10cm}
		\begin{verbatim}
		-if(engine.isContextSensitive() || 
		-"text".equals(buffer.getMode().getName()))
		+if(engine.isContextSensitive())
		\end{verbatim}
	\end{minipage} \\
\hline

Literal Negation Fault (LNF): Addition or removal of logical NOT (!) operator in fix atom which is being with a leading logical AND ($\&\&$)
	& \begin{minipage}[t]{10cm}
		\begin{verbatim}
		-if(ref != null && !ref.isPrepared())
		+if((ref != null) && ref.isPrepared())
		\end{verbatim}
	\end{minipage} \\
\hline

Term Negation Fault (TNF): Addition or removal of logical NOT (!) operator in fix atom which is being with a leading logical OR ($||$)
	& \begin{minipage}[t]{10cm}
		\begin{verbatim}
		-if(item == null || item.isStream())
		+if(item == null || !item.isStream())
		\end{verbatim}
	\end{minipage} \\
\hline

Expression Negation Fault (ENF): Addition or removal of logical NOT (!) operator in the whole fix expression
	& \begin{minipage}[t]{10cm}
		\begin{verbatim}
		-if(!row.isEmpty())
		+if(row.isEmpty())
		\end{verbatim}
	\end{minipage} \\
\hline

Operator Reference Fault (ORF): Change of logical operator/s without changing any atom in fix expression
	& \begin{minipage}[t]{10cm}
		\begin{verbatim}
		-if(palette != null || palette.length != 0)
		+if(palette != null && palette.length != 0)
		\end{verbatim}
	\end{minipage} \\
\hline

Multiple Fault: More than one fault scenario of nine fault classes are existing in the fix expression
	& \begin{minipage}[t]{10cm}
		\begin{verbatim}
		-if(config.top != null && config.top.length() != 0)
		+if(config == null)
		\end{verbatim}
	\end{minipage} \\
\hline
\end{tabular}
\label{table:ifcc:patterns}
%\vspace{-5pt}
\end{table*}

%table for LRF sub class definition ---------------------------
\begin{table*}[!ht]
\centering
\caption{LRF If-CC sub patterns and their corresponding examples }
%\vspace{-3pt}
%\begin{tabular}[c]{|p{4cm}|p{6cm}|l|l|}
\begin{tabular}[c]{|p{8cm}|p{9cm}|}
\hline
\textbf{LRF sub-patterns } & \textbf{Example} \\
\hline

Method Call Parameter Change (LRF-MCP): The bug atom and the fix atom are method call type and both contain the same method name but different parameter/s 
	& \begin{minipage}[t]{9cm}
		\begin{verbatim}
		-if(jEdit.getBooleanProperty("jdiff.horiz-scroll"))
		+if(jEdit.getBooleanProperty(HORIZ_SCROLL))
		\end{verbatim}
	\end{minipage} \\
\hline

Method Call Change (LRF-MCC): The bug atom and the fix atom are method call type but their method names are different
	& \begin{minipage}[t]{9cm}
		\begin{verbatim}
		-if(!dummyViews.contains(window))
		+if(!dummyViews.remove(window))
		\end{verbatim}
	\end{minipage} \\ 
\hline

Method Object Reference Change (LRF-MORC): The bug atom and the fix atom are method call type but they are different by their object references
	& \begin{minipage}[t]{9cm}
		\begin{verbatim}
		-if(data.getUsername() == null)
		+if(getUsername() == null)
		\end{verbatim}
	\end{minipage} \\ 
\hline

Relational Operator Change (LRF-ROC): The bug atom and fix atom use a different relational operator
	& \begin{minipage}[t]{9cm}
		\begin{verbatim}
		-if(index > 0  &&  index < this.buffers.size())
		+if(index >= 0 &&  index < this.buffers.size())
		\end{verbatim}
	\end{minipage} \\ 
\hline

Other Change (LRF-Other): Any change to an atom not fitting into the above four categories
	& \begin{minipage}[t]{9cm}
		\begin{verbatim}	
		-if(drag)
		+if(dragStart != null)
		\end{verbatim}
	\end{minipage} \\
\hline
\end{tabular}
\label{table:LRF:subpatterns}
%\vspace{-6pt}
\end{table*}

%table for non-fix patterns in if-cc fault ---------------------------
\begin{table*}[!ht]
\centering
\caption{Non-fix patterns (NF) in IF-CC and their corresponding examples }
%\vspace{-3pt}
%\begin{tabular}[c]{|p{4cm}|p{6cm}|l|l|}
\begin{tabular}[c]{|p{6cm}|p{11cm}|}
\hline
\textbf{Non-fix patterns} & \textbf{Example} \\
\hline

Exchange Operands in Equals Method (NF-EOEM): The operands of an equals method in bug and fix atoms are exchanged
	& \begin{minipage}[t]{11cm}
		\begin{verbatim}
		-if(methodName.equals("finalize"))
		+if("finalize".equals(methodName))
		\end{verbatim}
	\end{minipage} \\
\hline

Exchange Operands in an Infix Expression (NF-EOIE): The operands of an infix expressions in such a way as to have no semantic impact
	& \begin{minipage}[t]{11cm}
		\begin{verbatim}
		-if((-1 != row) && (!table.isRowSelected(row)))
		+if(row != -1 && !table.isRowSelected(row))
		\end{verbatim}
	\end{minipage} \\
\hline

Exchange Between Constants and Variables (NF-EBCV): Exchange between constant and variable in fix atom, where the variable always contains the constant value
	& \begin{minipage}[t]{11cm}
		\begin{verbatim}
		-if((slist == null) || (slist.getChildCount() != 3))
		+if((slist == null) || (slist.getChildCount() != BODY_SIZE))
		\end{verbatim}
	\end{minipage} \\
\hline

Variable Name Changed (NF-VNC):  Change the variable name in fix atom, where the values of the variables are always the same 
	& \begin{minipage}[t]{11cm}
		\begin{verbatim}
		-if(cell.getHeight() > maxHeight)
		+if(cell.getHeight() > currentMaxHeight)
		\end{verbatim}
	\end{minipage} \\
\hline

Addition/Removal of Meaningless Parenthesis (NF-ARMP): Addition/Removal of parenthesis in fix atom or expression, which do not change the order of evaluation of the expression
	& \begin{minipage}[t]{11cm}
		\begin{verbatim}
		-if(before < 0 || Character.isWhitespace(line.charAt(before)))
		+if((before < 0) || Character.isWhitespace(line.charAt(before)))
		\end{verbatim}
	\end{minipage} \\
\hline

Refactoring Changes (NF-RC): Addition or removal of redundant constant comparisons, or type casts 
	& \begin{minipage}[t]{11cm}
		\begin{verbatim}
		-if(mAutoFillEmptyCells == true)
		+if(mAutoFillEmptyCells)
		\end{verbatim}
	\end{minipage} \\
\hline
\end{tabular}
\label{table:nonfix:ifccpatterns}
%\vspace{-6pt}
\end{table*}

\normalsize

\section{Terminology}\label{Preliminaries}

%{\bf Fault/Bug:} A fault is defined in IEEE recommended practice~\cite{ieeestd:1633} as, (A) a defect in the code that can be the cause of one or more failures;  (B) an %accidental condition that causes a functional unit to fail to perform its required function.  A fault is synonymous with a bug.  A fault is an error that should be fixed with a %software design change.

{\bf Revision:} a set of changes to the software source code grouped by the developer.

%{\bf Version Control System (VCS):} software to keep track of  changes in project artifacts (primarily source code, but often including related documentation), 
%along with meta-information.  The changes are recorded when a revision is made, so that the state of the 
%source code after any revision can be recreated in the future.  In open source development, Concurrent Version System (CVS)~\cite{CVS:wiki} and Subversion(SVN)~\cite{SVN:wiki} %are widely used version control systems.

%{\bf Bug Tracking System (BTS):}  a system used to keep track of reports
%of software faults provided by developers, testers, and users.  These reports may include 
%descriptions of fault behaviour, evolving diagnoses, and ultimately proposed and adopted
%rectifications.  In open source development, Bugzilla~\cite{Bugzilla:web} is one of several
%widely used bug tracking systems.

{\bf Hunk:}  a single contiguous or near-contiguous section of source code which 
has undergone a change from revision to revision.  For our purposes, the division of 
the changes in a single revision into hunks is based on the rules of the Unix \emph{diff} program.  

{\bf Noise:}  a hunk which is not related to actually fixing a fault.  

{\bf Non fix hunk:}  a noise that reflects changes to programming code that do not 
reflect any semantic changes.
%is defined as a non fix hunk.

{\bf Atom:} a sub-condition in a single conditional expression.  For instance, a conditional expression is\verb+ x>0||y<10+; it consists two sub-conditions-\verb+ x>0+, and \verb+y<10+.  Here each sub-condition is called as an atom.

\section{Classification of IF-CC patterns} \label{Classification:if}
\subsection{IF-CC fix patterns}
Our intention is to examine whether existing Boolean expression testing methods are suitable for revealing common faults in real software systems, by assessing  the fault frequencies of fixed bugs in real software projects.  To collect this fault frequency information, we require a classification scheme which is automatically classifiable (to calculate the frequency), and aligned with the existing fault classes in the Boolean Expression fault (to select existing testing method).  We have devised a scheme which is adapted from the fault classes for Boolean expressions as described by Lau and Yu~\cite{LauYu:j-tosem-05}.  We have proposed ten IF-CC fix patterns:  Literal Reference Fault (LRF); Literal Omission Fault (LOF); Term Omission Fault (TOF); Literal Insertion Fault (LIF); Term Insertion Fault (TIF); Literal Negation Fault (LNF); Term Negation Fault (TNF); Expression Negation Fault (ENF); Operator Reference Fault (ORF); and Multiple Fault (MF).  
Table~\ref{table:ifcc:patterns} shows the definitions and examples of these fault classes.

In Boolean specification testing, truth values are considered for a single atom of the conditional.  However, a single atom in the conditional inside \verb+if+ statements may involve complex operations such as- arithmetic operations, method invocation, and comparison of objects.  This makes it more challenging to generate appropriate test data than if they are just simple Boolean variables.  Therefore, deeper knowledge about the fault of a single atom may help to generate suitable specification based test sets.  Preliminary investigation revealed LRF faults are more common.  We have therefore considered  Literal Reference Faults (LRF) in more detail and classified them into five subclasses.  The subclasses of LRF faults are Method Call Parameter Change (LRF-MCP), Method Call Change (LRF-MCC), Method Object Reference Change (LRF-MORC), Relational Operator Change (LRF-ROC), and Other Change (LRF:Other).  
Table~\ref{table:LRF:subpatterns} shows the definitions and examples of LRF sub-petterns.

%As our research aim was to examine whether existing boolean expression testing methods are suitable for revealing common faults in real
%software, we chose a classification scheme similar to that used by such methods.  Our scheme was therefore adapted from the fault
%classes for Boolean expressions as described by Lau and Yu~\cite{LauYu:j-tosem-05}.
%We propose ten types of the IF-CC fix patterns - namely, Literal Reference Fault (LRF), Literal Omission Fault (LOF), Term Omission Fault (TOF), Literal Insertion Fault (LIF), %Term Insertion Fault (TIF), Literal Negation Fault (LNF), Term Negation Fault (TNF), Expression Negation Fault (ENF), Operator Reference Fault (ORF) and Multiple Fault (MF).  
%Table~\ref{table:ifcc:patterns} shows the definitions of these fault classes with examples.

%%\subsection{LRF sub-patterns}

%The conditions inside \verb+if+ statements may involve arithmetic operations, method calls and other programming constructs.  Considering the difficulties of generating %specification based test sets, we have considered Literal Reference Faults (LRF) in more detail and classified them into five subclasses.  The subclasses of LRF faults are %Method Call Parameter Change (LRF-MCP), Method Call Change (LRF-MCC), Method Object Reference Change (LRF-MORC), Relational Operator Change (LRF-ROC), and Other Change %(LRF:Other).  
%Table~\ref{table:LRF:subpatterns} shows their definitions with examples.

\small
%\vspace{-3pt}
\begin{figure*}[!]  
\centering
  \includegraphics [scale = 0.65]{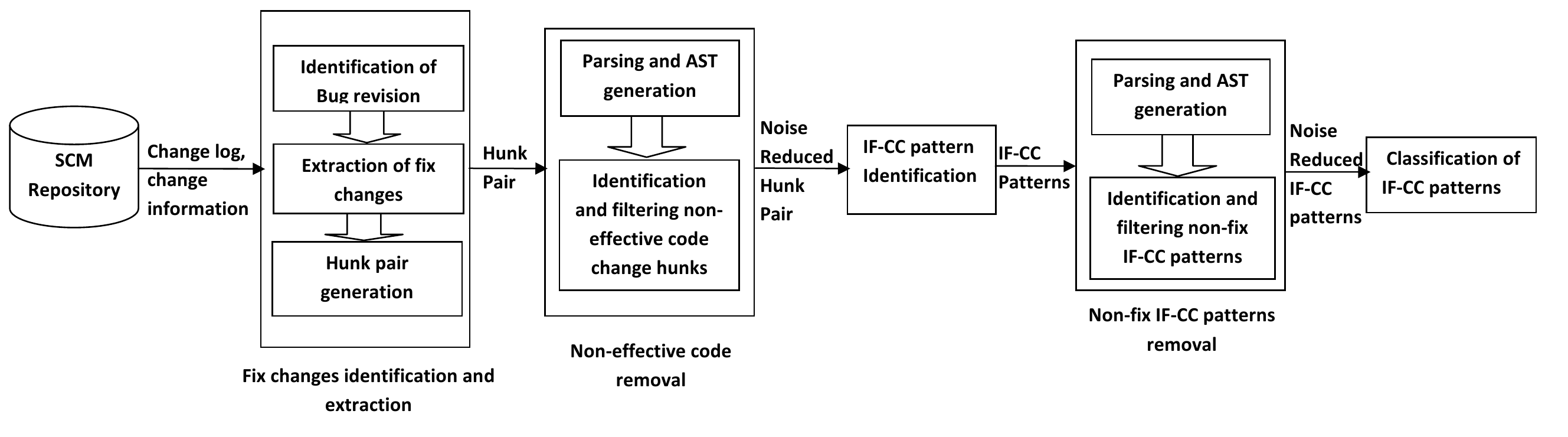}
  \vspace{-6pt}
\caption{Work-flow of identification, extraction and filtering of bug fix changes}
\label{WorkFlow}
\vspace{-4pt}
\end{figure*}

\normalsize
%\vspace{-4pt}

\subsection{IF-CC non-fix patterns} \label{Nonfix:if}
%We have observed that there are severa IF-CC change patterns which do not actually represent a change
%to program  contributing to make any semantic difference of the program.
%However, we defined the IF-CC fix patterns as the IF-CC non-fix patterns which indicate change necessarily with the programming language construct, but do not contribute in bug %fixing.  
Preliminary manual examination of change patterns showed that, similar to Jung \etal ~\cite{ClassJung} (see the result in ~\ref{manually:non:fix}), some common patterns of code changes to \verb+if+ conditionals which do not correspond to a code fix.  We found six such patterns: Exchange Operands in Equals Method (NF-EOEM), Exchange Operands in an Infix Expression (NF-EOIE), Exchange Between Constants and Variables (NF-EBCV), Variable Name Changed (NF-VNC), Addition/Removal of Meaningless Parenthesis (NF-ARMP), and Refactoring Changes (NF-RC).  These IF-CC non-fix patterns are described in Table~\ref{table:nonfix:ifccpatterns} with an example. These non-fix patterns will be identified and removed from further consideration just before the final classification of the IF-CC fix patterns.Please refer to Section~\ref{subsec:non-fix-reduction} for details.

\section{Research Methodology}\label{Research:Method}
Software configuration management (SCM) data is a valuable resource for empirical software engineering research.  For our study, we sought to  identify and extract bug fix changes from the repository.  Figure~\ref{WorkFlow} shows the workflow for identifying, extracting and filtering the bug fix changes. 

\subsection{Identification, extraction and hunk pair generation of bug fix changes}\label{Identification:Extraction:Filter}

\subsubsection{Identification of bug fixes}\label{keyword:extractor}

There are two ways to identify the bug fix revisions from the SCM repositories as described by data mining researchers.
\begin{itemize}
\item {\emph{Keyword searching approach:}} This approach for finding corrective maintenance from Version Control System (VCS)
%\footnote{Developers use version control system (VCS) to handle and store the changes on the code along with a log message for that changes. In open source development, Concurrent Version System (CVS) and Sub Version System (SVN) are widely used by the developers.} 
was first introduced by Mockus and Votta~\cite{Class8}.
Log messages written by developers accompanying revisions to the source code, stored by the version control system, are scanned for keywords (such as bug, fix, fault, bug ID, and so on).  The data thus collected is called VCS data.  This approach has risks of including both false positives and false negatives.
False negatives occur when developers do not include the keywords in their log messages when fixing bugs.  False positives can occur when, for instance, 
unrelated changes to source code are also included in a revision that contains a bug fix. 

\item {\emph{Linking approach: }} This approach was proposed by Fisher \etal~\cite{Class9}.  
It establishes a link between bug tracking system (BTS) data and VCS data so as to get more accurate bug data from the available repository information.
%\footnote{bug tracking system (BTS) is used to keep track of bug and its information. So that the bugs could be handled in more finger grained way.  In open source development, Bugzilla and Bug tracker are widely used by the developers.}
In a bug tracking system, all issues (not all which are bugs) are assigned a unique numeric identifier.  In the linking approach, the VCS data is searched for issue ID to find the code changes corresponding to the issue.  However, the links are imperfect; Ayari \etal~\cite{Class4} found that there were many missing links between the Bugzilla (BTS) and CVS (VCS) repositories for the Mozilla project.  Only a fraction of BTS entries are actually bugs; Ayari \etal\ showed, by manual inspection of a subset of bug reports that only 58\% issues in the BTS repository actually indicated corrective maintenance.
% \end{list}
\end{itemize}
In this paper, we chose to use the keyword searching approach to identify the bug fix revisions due to the low identification rate of the linking approach.

\subsubsection{Extraction of bug fix changes}\label{Extraction:bug:hunks}

Bug fix changes were extracted from the SVN repository using the {\tt svn diff} command.  
Developers normally change different sections of a single or multiple files to fix a bug.  The
resulting ``diffs'' show the differences between the two revisions, showing different sections in a single file with a separator containing with the range information in Unidiff format~\cite{Unidiff:web}.  For our purposes, each distinct section identified by diff was identified as a separate bug fix hunk that represents the bug fix change.

\subsubsection{Hunk pair generation}
Each bug fix hunk was then extracted out of the diff file.  Hunk pairs are generated for each hunk with the help of the {\tt patch}, applying the fix hunk to the previous revision of the file.  These hunk pairs were described as the ``bug hunk'' and ``fix hunk''.
 
%This fix hunks are collected from the Version Control System (VCS) using extracted fix revision number. ** hunk pair generation

%\subsubsection {Noise~\footnote{he change hunk which is not related to programming language constructs or which is not contributing on bug fixing, is defined as {\tt noise}} filtering} \label{Noise:filtering}

\subsection{Non-effective hunk filtering}\label{Noise:filtering}

Removing noise - change hunks that are not actually code changes related to bug fixing - takes places at two stages in our workflow.   The first 
class of noises filtered in our approach is ``non-effective change hunks''.  These include changes that have no effect on the source code as perceived
by the compiler - for instance, changes to comments.  They are removed from further processing immediately after hunk pair generation, as shown in Figure~\ref{WorkFlow}.  

To remove these hunks, we generate the parse trees including only the effective Java codes of the IF-CC hunk pairs (excluding, for instance, comments). If the ``effective'' trees for the code segments are same, we identify the hunk as non-effective and exclude it as a noise.

Elimination of effective non-fix hunks was done after identification of IF-CC patterns and is described in section~\ref{subsec:non-fix-reduction}.

%we eliminate those that fall into the non-fix bug patterns as described in Table~\ref{table:nonfix:ifccpatterns}.  As we did not eliminate some of the non-fix hunks from our data set, it would affect the result by adding some false positive data.  \fbox{Shimul and Robert - please check the logic of the sentence?} \fbox{If we know that it is a non-fix hunk, we should remove it.} \fbox{Why keep it?}Those non-fix hunks which we did not eliminate will be classified as LRF:other.
% Should discuss this in results? As such inclusion of some non-fix hunks into \emph{LRF:other} category, we believe it will not affect, largely, on our conclusion.

\subsection{Analysis and pattern classification}

Fluri \etal~\cite{10.1109/TSE.2007.70731} proposed an algorithm to identify particular changes between two versions of a program. It compares every node of two ASTs (Abstract Syntax Tree) of the two revisions of a file to identify specific changes.  As such two ASTs are to be generated for every hunk pair,  and each node of the ASTs are to be compared to identify particular change in a fix hunk.  In this study we adopt their approach to identify the IF-CC fix patterns.

\subsubsection{Identification of IF-CC}\label{Identification:IF-CC}
The two generated ASTs are compared node by node.
While any node difference is found and both nodes are \verb+if+ statements, our tool then considers whether the change is actually an \verb+if+ conditional change.  Only the following cases were considered as a hunk contained an IF-CC pattern.
%\begin{list}{$\bullet$}{\setlength{\leftmargin}{0.05in}}
\begin{itemize}
	\item While the \emph{conditions} of the two \verb+if+ statements are not the same but their \verb+then+ statements are the same, this fix hunk is considered as IF-CC fix pattern.
	\item While the \emph{conditions} of the two \verb+if+ statements are not the same and their \verb+then+ statements are not the same, this fix hunk can be considered or rejected as IF-CC fix pattern as follows - 
	
(i) If the \verb+then+ block  was more than 5 statements long, and no more than two of these differed, it was considered as an IF-CC hunk pattern.

(ii) If the \verb+then+ block was less than 5 statements and the differed statement/s are differed due to refactoring (such as a variable renaming, or changing a casting keyword), it was considered as IF-CC hunk pattern.

(iii) All other cases were not considered IF-CC hunk patterns and these were ignored for further analysis.
\end{itemize}
%
%\fbox{Shimul and Robert: I do not understand this.} 
%\fbox{For both (i) and (ii) above, we always classify them as}
%\fbox{IF-CC hunk pattern. There is no second option as suggested above!}

%\fbox{Shimul and Robert - The two paragraphs above are similar}
%\fbox{as well as contradicting. Similar - all mentioned about using JDT}
%\fbox{Contradicting - manually verify two artifacts; all seven?}
\subsubsection{Non-fix IF-CC reduction techniques}
\label{subsec:non-fix-reduction}

The second group of ``noise hunks'' are known as ``Non-fix change hunk'', which are code changes performed at the same time as a bug fix, which do affect 
the code's parse tree, but are not actually related to the bug fix. For instance, the addition of parentheses to an expression
for readability.   Non-fix IF-CC hunks were identified and eliminated, where this was possible.  However, not all such non-fix hunks can be straightforwardly
removed.

To perform this elimination, a case-by-case check was performed on each identified IF-CC change hunk to see whether the parse trees match patterns indicating a non-fix hunk.  For instance, consider a case when a condition of an {\tt if} is an {\tt equals} method and the condition is changed in the fix revision by exchanging the operands of the {\tt equals} method, leaving everything else unchanged.  This is a non-fix change, as the change has no semantic impact on the program.  For example, consider a case where \verb+if(methodName.equals("finalize"))+ is changed to \verb+if("finalize".equals(methodName))+.  Our tool's static analyzer checks whether both conditions are {\tt equals} methods.  If they are both {\tt equals} methods, it compares the operand set of the {\tt equals} method of the bug revision with the operands set of the {\tt equals} method of the fix revision.  In the example, the operand set for bug revision is \{\verb+methodName+ , \verb+"finalize"+\} and the operand set for the fix revision is \{\verb+"finalize"+, \verb+methodName+ \}.  As the two operand sets are same, it is considered as NF-EOEM (Exchange Operands in Equals Method) and eliminated as noise.  Among six identified non-fix IF-CC hunks (See in Section~\ref{Nonfix:if}), NF-EOEM, NF-EOIE, and NF-ARMP are eliminated as they are detected with static analysis.  Other non-fix IF-CC patterns are detectable using some complex analysis techniques (such as- class analysis), which we did not implement in this work.  The undetected non-fix IF-CC patterns affect the result by adding some false positive data.  Those non-fix hunks which we did not eliminate were classified as LRF:other, thus inflating the frequency of this pattern by some amount.  A manual check on a subset of data was conducted to determine the likely impact of this and the results are reported in Section~\ref{manually:non:fix}.

\subsubsection{IF-CC Pattern classification}

The filtered IF-CC change hunks were then classified into the categories described in Table~\ref{Classification:if}.  We used Fluri \etal's~\cite{10.1109/TSE.2007.70731} tree traversal algorithm as the basis of our classifier.  The classifier compared each atom of the if condition
in turn and used a simple decision tree approach to determine which pattern was applicable.  

\section{Tools Implementation and verification} \label{tools:verification}
\subsection{Extractor}
We have developed a module in Perl to extract the bug fix hunks automatically.  It uses a two-step process:
%\begin{list}{$\bullet$}{\setlength{\leftmargin}{0.05in}}
\begin{itemize}
\item \textbf{Identify the fix revision:}
The keyword searching approach described in Section~\ref{keyword:extractor}, is used to extract the bug fix revisions.  Following the general approach of (for example) Ayari~\etal~\cite{Class4}, we used the following regular expression in our extractor to extract fix revisions: \\
%\vspace{-4pt}
\begin{quote}
\vspace{-6pt}
\emph{fix(e[ds])?$|$bugs?$|$patches?$|$defects?$|$faults?}
\end{quote}
\vspace{3pt}
		
\item \textbf{Extract fix hunk: }
Our tool extracted the change for a specific bug fix revision using the \emph{svn diff} command and the identified revision number.  From this change information we have separated different regions of changes as different hunks. 
%\end{list}
\end{itemize}
After extracting the fix hunks, hunk pairs are generated using the Unix \emph{patch} command using hunks and their corresponding revision file.  To verify the output of the extractor module, we manually checked all the hunk pairs of \emph{Checkstyle}, one of the seven artifacts studied in this study.

\subsection{Analyser}

We have developed an analyser module in Java that makes use of the Eclipse JDT Tooling~\cite{JDT:web} to identify the IF-CC fix patterns.  The module takes a hunk pair as input and generates two ASTs.  These two ASTs are compared node-to-node to see whether differed nodes in two ASTs are both \verb+if+ condition or not.  While the differed nodes are \verb+if+ condition and the \emph{condition} of two if statements are not same but \emph{then} statements are same, this fix hunk is extracted as IF-CC fix pattern.  Other differed nodes with \verb+if+ conditions changes were flagged and checked manually to determine whether they were IF-CC hunks.  To verify the correctness of the Analyser, we manually checked all the identified IF-CC patterns of \emph{Checkstyle} and \emph{JabRef} by the analyser.
     
%\subsection{IF-CC Fix Pattern Classifier}

\subsection{Classifier}
We have developed a classifier module in Java, again utilizing the Eclipse JDT Tooling, for classifying the IF-CC fix patterns identified by the Analyser according to the described fault classes in Section~\ref{Classification:if}.  To verify the correctness of the Classifier, we have manually verified all the classified IF-CC patterns of all seven artifacts.

\section{Results}\label{Result}

%\subsection {Data set}
%We have selected the software projects as data set which have a reasonable usability and maintainability history.  We have selected all the open source software because we have easy access of their repositories.  Most of the software projects in our data set were using CVS (Concurrent Version Control System) repositories early in their development, but they later migrated to  SVN (Subversion).  In SVN, a revision is defined as a single commit, which may contain changes to a single file or multiple files.  We have considered only the SVN (Subversion) repository data for our research purpose.  The selected software projects and their maintenance histories are described in the table~\ref{table:history}.

\subsection{Software Artifacts}

We have selected seven open source systems implemented in Java for our study.  All seven artifacts have a long term maintenance history.  We have considered change histories ranging from 4 years to 10 years, depending on the available revisions in  {\tt SVN} repositories.  We selected open source software projects because we have easy access of their repositories.  We have extracted the bugs from the incremental revisions of the software, not just the bugs found in  customer releases.  As unit testing methods are often applied to incremental versions as well as major ones, we believe it is appropriate to consider bugs in incremental revisions as well as major releases.  Moreover, it also provides a much larger sample to work with.

%\small

%table Analysed project description
%\begin{landscape}
\begin{table*}[!t]
\centering
%\vspace{-2pt}
\caption{Analysed Projects and their change descriptions}
%\vspace{-3pt}
\begin{tabular}[c]{lccccccc}
\hline\hline
%\multirow{2}{*}{
{\minitab[l] {Project\\  Name}} & {\minitab[l] {Considered\\  Period}} & {\minitab[l]{No. of\\Revisions}} & {\minitab[l]{No. of fix\\ Revisions}} 
& {\minitab[l]{ Total\\  hunks}}
& {\minitab[l]{Effective\\  hunks}}
& {\minitab[l]{IF-CC\\  hunks}}
& {\minitab[l]{\% of\\  IF-CC}}\\

\hline
Checkstyle&20/06/2001-15/08/2010& 2538 & 609 & 3404 & 1685 &92&5.5\%\\
JabRef&14/10/2003-21/08/2010 & 3316 & 413& 4131 & 2032&106&5.2\%\\
iText &28/11/2000-15/08/2010& 4576 & 506 & 2813 & 1063&108&10.2\%\\
%Columba & -& 456 & 189\\
DrJava &19/06/2001-30/08/2010 & 5419 & 1208 & 38014 & 11781 &511&4.3\%\\
jEdit &01/07/2006-17/08/2010& 18382 & 3767 & 34651 & 19904 &1027&5.2\%\\

PMD &21/06/2002-22/09/2010& 7141& 988 & 10186 & 4444 &475&10.7\%\\
%MegaMek &-& 7717 & 831\\
FreeCol &14/01/2002-30/09/2010&7484&1431&13571 &3226 &215&6.7\%\\

\hline
\end{tabular}
\label{table:history}
%\vspace{-3pt}
\end{table*}

\normalsize

%
%% table for bug if-cc extractions and their frequencies over different projects 
%\begin{table}[!h]
%\centering
%\caption{Extracted IF-CC hunks and their frequencies over seven projects}
%\begin{tabular}[c]{lccc}
%\hline\hline
%{\minitab[l] {Project\\  Name}}& {\minitab[l]{Analysed\\		hunks}}& {\minitab[l]{IF-CC\\ hunks}} &{\minitab[l]{\% of \\IF-CC}} \\
%\hline
%Checkstyle&1685&92&5.5\%\\
%JabRef&2032&106&5.2\%\\
%iText&1063&108&10.2\%\\
%DrJava&11781&511&4.3\%\\
%jEdit&19904&1027&5.2\%\\
%PMD&4444&475&10.7\%\\
%FreeCol&3226&215&6.7\%\\
%\hline
%\end{tabular}
%\label{table:iffriquency}
%\end{table}

%\small
%table for frequencies of different if-cc faults
\begin{table*}[!]
\scriptsize
\centering
%\vspace{-2pt}
\caption{Classified If-CC patterns in different fault categories and their frequencies}
%\vspace{-3pt}
%\begin{tabular}[c]{|p{4cm}|p{6cm}|l|l|}
\begin{tabular}[c]{|p{1.8cm}|p{.69cm}|p{.69cm}|p{.69cm}|p{.69cm}|p{.69cm}|p{.69cm}|p{.69cm}|p{.69cm}|p{.69cm}|p{.69cm}|p{.69cm}|p{.69cm}|p{.69cm}|p{.69cm}|}
\hline
\textbf{Fault category} & \multicolumn{2}{|c|}{\begin{turn}{0}\textbf{Checkstyle}\end{turn}} & \multicolumn{2}{|c|}{\begin{turn}{0}\textbf{JabRef}\end{turn}} & \multicolumn{2}{|c|}{\begin{turn}{0}\textbf{iText}\end{turn}} & \multicolumn{2}{|c|}{\begin{turn}{0}\textbf{DrJava}\end{turn}}&\multicolumn{2}{|c|}{\begin{turn}{0}\textbf{jEdit}\end{turn}}&\multicolumn{2}{|c|}{\begin{turn}{0}\textbf{PMD}\end{turn}}&\multicolumn{2}{|c|}{\begin{turn}{0}\textbf{FreeCol}\end{turn}}\\
\hline

LRF:MCC &6&7.1\%&4&3.8\%&3&2.8\%&164&32.8\%&159&15.7\%&274&58.7\%&7&3.3\%\\
LRF:MCP&5&5.9\%&7&6.7\%&4&3.7\%&29&5.8\%&67&6.6\%&44&9.4\%&14&6.5\%\\
LRF:MORC &3&3.5\%&19&18.1\%&4&3.7\%&37&7.4\%&34&3.4\%&7&1.5\%&5&2.3\%\\
LRF:ROC&4&4.7\%&5&4.8\%&4&3.7\%&6&1.2\%&24&2.4\%&2&0.4\%&6&2.8\%\\
LRF:Other &26&30.6\%&23&21.9\%&50&46.3\%&137&27.4\%&295&29.1\%&67&14.3\%&43&20.1\%\\

{\textbf{Category total}}&{\textbf 44}&{\textbf 51.8\%}&{\textbf 58}&{\textbf 55.2\%}&{\textbf 65}&{\textbf 60.2\%}&{\textbf 373}&{\textbf 74.6\%}&{\textbf 579}&{\textbf 57.1\%}&{\textbf 394}&{\textbf 84.4\%}&{\textbf 75}&{\textbf 35.0\%}\\
\hline
\textbf{LOF}&18&21.2\%&19&18.1\%&16&14.8\%&51&10.2\%&164&16.2\%&18&3.9\%&54&25.2\%\\
\hline
\textbf{TOF}&10&11.8\%&13&12.4\%&6&5.6\%&25&5.0\%&108&10.7\%&28&6.0\%&30&14.0\%\\
\hline
\textbf{LIF}&0&0.0\%&0&0.0\%&2&1.9\%&10&2.0\%&41&4.0\%&1&0.2\%&6&2.8\%\\
\hline
\textbf{TIF}&1&1.2\%&3&2.9\%&3&2.8\%&3&0.6\%&22&2.2\%&1&0.2\%&3&1.4\%\\
\hline
\textbf{LNF}&0&0.0\%&0&0.0\%&0&0.0\%&3&0.6\%&0&0.0\%&1&0.2\%&0&0.0\%\\
\hline
\textbf{TNF}&0&0.0\%&0&0.0\%&1&0.9\%&0&0.0\%&0&0.0\%&0&0.0\%&1&0.5\%\\
\hline 
\textbf{ENF}&0&0.0\%&1&1.0\%&1&0.9\%&5&1.0\%&8&0.8\%&0&0.0\%&4&1.9\%\\
\hline
\textbf{ORF}&0&0.0\%&1&1.0\%&2&1.9\%&0&0.0\%&3&0.3\%&1&0.2\%&0&0.0\%\\
\hline
%\multirow{7}{*}{\textbf{Double fault}} & LRF&11&1&0&28\\
%&TOF&1&6&2&2\\
%&LIF&5&0&0&7\\
%&TIF&4&0&0&2\\
%&ROF&1&0&0&0\\
%&LOF&0&3&0&9\\
%&{\textbf{Category Total}}&(13.4\%)   11&(7.5\%)   5&(3.4\%)   1&(8.0\%)  24\\
%\hline
{\textbf{Multiple}}&12&14.1\%&10&9.5\%&12&11.1\%&30&6.0\%&89&8.8\%&23&4.9\%&41&19.2\%\\
\hline
\hline
%{\textbf {Analysed Fault}}&\multicolumn{2}{|c|}{&&&&&&&&&&\\
\textbf{Analysed Fault} & \multicolumn{2}{|c|}{92} & \multicolumn{2}{|c|}{106} & \multicolumn{2}{|c|}{108} & \multicolumn{2}{|c|}{511}&\multicolumn{2}{|c|}{1027}&\multicolumn{2}{|c|}{475}&\multicolumn{2}{|c|}{215}\\
\textbf{Total Fault} & \multicolumn{2}{|c|}{85} & \multicolumn{2}{|c|}{105} & \multicolumn{2}{|c|}{108} & \multicolumn{2}{|c|}{500}&\multicolumn{2}{|c|}{1014}&\multicolumn{2}{|c|}{467}&\multicolumn{2}{|c|}{214}\\
\textbf{Non-fix} & \multicolumn{2}{|c|}{7} & \multicolumn{2}{|c|}{1} & \multicolumn{2}{|c|}{0} & \multicolumn{2}{|c|}{11}&\multicolumn{2}{|c|}{13}&\multicolumn{2}{|c|}{8}&\multicolumn{2}{|c|}{1}\\

\hline
\end{tabular}

\label{table:frequency:if:fault}
%\vspace{-3pt}
\end{table*}
%\end{landscape}

\normalsize

\subsection{Artifact properties and extracted bug fix data}\label{hunk:extractor}
 Table~\ref{table:history} shows the considered maintenance histories, the number of revisions, extracted bug fix revisions, extracted fix hunks, effective fix hunks and extracted IF-CC hunks for the software artifacts have found in their corresponding {\tt SVN} repositories.  The result suggests that neither the rates of revisions over time nor the rates of extracted bug fix hunks are consistent in different software artifacts.  This should be kept in mind when considering cross-project comparisons.       

We have analysed the effective fix hunks through our analyser, and identified the IF-CC fix hunks.  Table~\ref{table:history} shows that the IF-CC fix hunk frequencies varied from 4.3\%, in \emph{DrJava} to 10.7\%, in \emph{PMD}.  Despite the variation in relative frequency in each artifact, we had sufficient examples of IF-CC fix patterns to collect meaningful statistics about their properties for all seven artifacts.

\subsection{Frequencies of different IF-CC fix patterns}\label{s:freq}

Table~\ref{table:frequency:if:fault} shows the numbers and frequencies of different IF-CC fix patterns along with the detected non-fix patterns across the seven software artifacts.  It shows that the percentage of different fault classes in seven software projects are different.  However, it is clear that the rank of frequencies of different fault classes are quite consistent over the seven artifacts.  

LRF is the most common fault class in all the projects, although its frequency varies from 35.0\% in \emph{FreeCol} to 84.4\% in \emph{PMD}.  The mean frequency for LRF is more than 50\% of total IF-CC faults.  In other words, over half of all \verb+if+ condition faults are related to a single condition inside the \verb+if+ expression.  The second and third largest frequency groups are LOF and TOF respectively in all the artifacts, except in \emph{PMD} where their rankings were reverse.  Thus almost one-third of the \verb+if+ condition faults are due to omission of subconditions inside the \verb+if+ expression.  The other fault classes (LIF, TIF, LNF, TNF, ENF, ORF) had frequencies of between 0 to 4\% in the different software projects, which indicates that these fault classes are much less frequent than the major three class of faults.  Besides those ``single fault'' classes, a considerable proportion of all faults are \emph{multiple faults}, representing between 4.9\% to 19.2\% of all faults across all the software artifacts.  Multiple faults are the most next common fault type with the \verb+if+ expression after LRF, LOF and TOF.

%Unlike other research, we classify multiple faults as multiple occurrences of the single faults and \emph{complex faults} are faults that involving complicated changes of the %\verb+if+ expressions as well as other single faults. Complex faults are the most next common fault type with the \verb+if+ expression after LRF, LOF and TOF.  
%\fbox{Shimul - Is there any data re complex faults?}

From Table~\ref{table:frequency:if:fault}, the LRF subcategories vary substantially from project to project.  Among them, either the LRF:Other or the LRF:MCC are the major LRF sub-classes in the seven projects.  The LRF:MCP and LRF:MORC represent 3.7\% to 9.4\% and 1.5\% to 18.1\% respectively of total faults related to IF-CC patterns.  LRF:ROC faults were significant in some projects but not others, with their frequency varying between 0.4\% to 4.8\% of all IF-CC faults.  
 
%Our IF-CC fix pattern classifier have classified the IF-CC fix patterns and also identified some of the non-fix hunks.  In Table~\ref{table:frequency:if:fault}, we have described the frequencies of different fault classes and number of identified non-fix hunks of IF-CC fix patterns over different software artifacts.  Inspite of  the \emph{LRF} is the major group of faults in all the software artifacts    

Some of the non-fix patterns that described in Section~\ref{Nonfix:if} are detected by our tool and have been summarized in the Table~\ref{table:frequency:if:fault}.  The results suggest that non-fix pattern detection rates are not the same across different projects.  Perhaps, neither all the non-fix patterns are present nor the frequencies are the same in all the software artifacts.  As a result, some of the non-detected non-fix IF-CC patterns in our result are considered and are classified as LRF:Other.   

%\small
% table for non-fix frequencies
\begin{table}[!h]
\centering
%\vspace{-2pt}
\caption{Frequencies of non-fix IF-CC patterns in our sample set}
%\vspace{-3pt}
\begin{tabular}[c]{lccc}
\hline\hline
{\minitab[l]{Project\\Name}}& {\minitab[l]{Total\\Hunk}}& {\minitab[l]{Non-fix\\ Hunk}} &{\minitab[l]{Our tool\\Identified}} \\
\hline
Checkstyle & 92 & 13 (14.1\%) &7\\
JabRef & 106 & 5 (4.7\%) &0\\
iText & 108 & 5 (4.6\%) & 0\\
DrJava & 100 & 9 (9.0\%) & 3\\
jEdit & 110 & 3 (2.7\%) & 2\\
{\bf Total} &{\bf 516} & {\bf 35 (6.8\%)} & {\bf 12}\\
\hline
\end{tabular}
\label{nf:frequency}
%\vspace{-6pt}
\end{table}
  
%\vspace{-2pt}
\normalsize

\subsection{Manual study to trace non-fix hunks} \label{manually:non:fix}

%\subsection{Noise Reduction}
The noises with the non-code changes described in~\ref{Noise:filtering} are removed from our data.  
In a previous general study of fault pattern data, Jung \etal ~\cite{ClassJung} identified  11 common non-fix patterns, which represented almost 8.3\% of the total hunks in their manual inspection.  However, none of the 11 non-fix patterns related to if-condition changes.  While this gave us some indication that
most IF-CC fix hunks did indeed represent effective code changes, we still considered it prudent to conduct a manual check a representative set of IF-CC fix hunks, to identify and classify non-fix IF-CC patterns (if they exist).  All extracted IF-CC patterns of \emph{Checkstyle}, \emph{JabRef}, \emph{iText} and randomly selected sets from other two artifacts (\emph{DrJava} and \emph{jEdit}) were inspected manually for this study.  We have analysed a total 516 IF-CC fix patterns selected from five software systems (\emph{Checkstyle}, \emph{JabRef}, \emph{iText}, \emph{DrJava} and \emph{jEdit}) and found 35 non-fix IF-CC patterns, which is almost 6.8\% of the analysed IF-CC hunks (see in Table~\ref{nf:frequency}).  We have classified them into six categories which is described in Section~\ref{Nonfix:if}.

%\small
% table for non-fix hunk data
\begin{table}[t]
\centering
%\vspace{-2pt}
\caption{Non-fix IF-CC patterns found in our sample set}
%\vspace{-3pt}
\begin{tabular}[c]{|c|rrrrr|r|}
\hline
{\minitab[l]{Non-fix IF-CC\\ patterns}} & \begin{turn}{90}Checkstyle\end{turn} &\begin{turn}{90}JabRef\end{turn} &\begin{turn}{90}iText\end{turn} &\begin{turn}{90} DrJava\end{turn} &\begin{turn}{90}jEdit\end{turn}& \begin{turn}{90}{\bf Total}\end{turn}\\
\hline
NF-EOEM & 5 & 0 & 0 & 0 & 1 &{\bf 6}\\
NF-EOIE & 0 & 0 & 0 & 0 & 1 &{\bf 1}\\
NF-EBCV & 3 & 0 & 0 & 0 & 0 &{\bf 3}\\
NF-VNC & 3 & 3 & 4 & 6 & 1 & {\bf 17}\\
NF-ARMP & 2 & 2 & 0 & 3 & 0 & {\bf 5}\\
NF-RC & 0 & 5 & 1 & 0 & 0 & {\bf 3}\\
\hline
{\bf Total} & {\bf 13} & {\bf 5} & {\bf 5} & {\bf 9} & {\bf 3} & {\bf 35}\\

%{\bf Total} \\
\hline
\end{tabular}
\label{nf:sample}
%\vspace{-6pt}
\end{table}
%\vspace{-3pt}
\normalsize

Table~\ref{nf:sample} shows that the frequencies of six non-fix IF-CC patterns are not similar in different software artifacts.  Only one non-fix pattern, Variable Name Changed (NF-VNC), among six was found in all software artifacts.  The NF-VNC was the most common non-fix pattern seen, representing almost half of the total non-fix patterns observed.  The frequencies of IF-CC non-fix patterns in five software artifacts varies from 2.7\% to 14.1\%, which indicates their inconsistent appearance in different software artifacts (see Table~\ref{nf:frequency}).  Perhaps, this would happen due to different software applications having different code complexity or, different programmers choosing different coding styles.  For instance, some programmers use redundant parentheses to increase the code readability whereas some programmers do not.

\section {Discussion}\label{Discussion}

%\small
%table for spearman correlation for if-cc patterns
\begin {table*}[!]
%\scriptsize
\centering
%\vspace{-2pt}
\caption{Spearman rank correlation values of IF-CC for seven artifacts}
%\vspace{-3pt}
\begin{tabular}{cccccccl}
%\begin{tabular}{p{0.75cm}p{0.57cm}p{0.57cm}p{0.57cm}p{0.57cm}p{0.57cm}p{0.57cm}l}
\hline
Checkstyle&JabRef&iText&DrJava &jEdit&PMD&FreeCol&\\
%&\begin{turn}{90}Checkstyle\end{turn}&\begin{turn}{90}JabRef\end{turn}&\begin{turn}{90}iText\end{turn}&\begin{turn}{90}DrJava\end{turn}&\begin{turn}{90}jEdit\end{turn}&\begin{turn}{90}PMD\end{turn}&\begin{turn}{90}FreeCol\end{turn}\\
%\multicolumn{2}{c}{MegaMek}\\
\hline
1.00&0.94&0.94&0.85&0.89&0.86&0.86&Checkstyle \\
&1.00&0.92&0.78&0.88&0.81&0.80&JabRef\\
&&1.00&0.81&0.93&0.85&0.87&iText \\
&&0.81&1.00&0.93&0.79&0.96&DrJava\\
&&&&1.00&0.84&0.95&jEdit\\
&&&&&1.00&0.73&PMD\\
&&&&&& 1.00&FreeCol\\
\hline

\end{tabular}

\label{Spearman:rank:ifcc}
%\vspace{-4pt}
\end {table*}
%\vspace{-3pt}
\normalsize

%table for spearman correlation for LRF sub-patterns
\begin {table*}[!]
%\scriptsize
\centering
%\vspace{-2pt}
\caption{Spearman rank correlation values of LRF sub patterns for seven artifacts}
%\vspace{-3pt}
\begin{tabular}{cccccccl}
%\begin{tabular}{p{0.75cm}p{0.57cm}p{0.57cm}p{0.57cm}p{0.57cm}p{0.57cm}p{0.57cm}l}
\hline
Checkstyle&JabRef&iText&DrJava&jEdit&PMD&FreeCol&\\
%&\begin{turn}{90}Checkstyle\end{turn}&\begin{turn}{90}JabRef\end{turn}&\begin{turn}{90}iText\end{turn}&\begin{turn}{90}DrJava\end{turn}&\begin{turn}{90}jEdit\end{turn}&\begin{turn}{90}PMD\end{turn}&\begin{turn}{90}FreeCol\end{turn}\\
%\multicolumn{2}{c}{MegaMek}\\
\hline
 1.00&0.10&0.22&0.60&0.90&0.80&0.90&Checkstyle\\
 &1.00&0.89&0.00&0.30&-0.10&0.30&JabRef\\
 &&1.00&-0.22&0.22&-0.22&0.45&iText\\
 &&&1.00&0.80&0.90&0.30&DrJava\\
 &&&&1.00&0.90&0.80&jEdit\\
 &&&&&1.00&0.60&PMD\\
 &&&&&& 1.00&FreeCol\\
\hline

\end{tabular}

\label{Spearman:rank:lrf}
%\vspace{-3pt}
\end {table*}
%\vspace{-3pt}
\normalsize

\subsection{Cross project frequency similarity }\label{Cross:project}
We found that most of the IF-CC fix patterns can be classified based on the fault classes of specification based testing of Boolean expressions, which is the answer of our first research question (RQ1).  To answer the second research question (RQ2), we have calculated the Spearman's rank correlation coefficient~\cite{Spearman:Wiki} of ten major IF-CC fix patterns for all the seven projects (see Table~\ref{Spearman:rank:ifcc}) and found most of the correlation values are higher than 0.8 (except two values).  The statistical significance of these 21 correlation values are calculated using the Holm-Bonferroni method~\cite{holm-bonferroni}, with an overall $\alpha =0.05$.  We have found all 21 correlations are statistically significant.  We can conclude that the ranking of IF-CC fix patterns is truly similar 
across projects.

We also calculated the Spearman correlation coefficients for the sub-patterns of LRF across the seven software artifacts (see Table~\ref{Spearman:rank:lrf}).  The statistical significance of these correlations are calculated using Holm-Bonferroni method~\cite{holm-bonferroni}, and found that the correlations were not 
significant with an overall $\alpha =0.05$.  These result suggests that the appearance of LRF subpatterns are not consistent over projects.

%\subsection{Effectiveness of Boolean Expression testing in revealing if conditional faults}
\subsection{Implications for Testing}

As discussed in Section~\ref{s:freq}, the  three most common fault categories are
\begin{enumerate}
\item LRF ranging from 35.0\% to 84.4\%
\item LOF ranging from 3.9\% to 25.2\%
\item TOF ranging from 5.6\% to 14.0\%
\end{enumerate}
In the seven artifacts, the fraction of all if-condition faults falling in to these categories ranges from 74.2\% (FreeCol) to 94.3\% (PMD), which can be calculated from Table~\ref{table:frequency:if:fault}.
Hence, any testing methodologies which are good at detecting LRF, LOF and TOF (for example, the MUMCUT testing techniques~\cite{YuLauChen:j-jss-06-1}) would be very useful in revealing almost $\frac{3}{4}$ of the IF-CC bug fixes.
Hence, if software testers are to perform testing given limited time and resources, our results suggest that selecting those testing techniques that can reveal LRF, LOF and TOF would be a reasonable choice.  The high correlation of fault category frequency across the seven artifacts (as discussed in Section~\ref{Cross:project}), therefore, suggests that this recommendation is generally applicable of testing \verb+if+ conditionals across different types of Java software systems.

Furthermore, as suggested by the fault class hierarchy~\cite{LauYu:j-tosem-05}, test cases that can detect those at the lower part of the hierarchy (e.g. LRF, LOF and TOF) will be able to detect those corresponding faults on the upper part of the hierarchy (e.g. TNF, LNF, ORF and ENF).  
Hence, these test cases have a good chance to detect faults categorized as LNF, TNF, ORF and ENF.

%\vspace{-2pt}
\section{Threats to validity}\label{Threat:validity}
%\vspace{-2pt}
\subsection{Internal Validity}

The \emph{Extractor} identified bug fix revisions using the keyword searching approach (described in section~\ref{keyword:extractor}). Developer omission of
these keywords in log messages, or the inclusion of unrelated changes in a revision tagged as a bug fix, will result in false negative and false positives
respectively.  False negatives are a less significant concern, as there is no obvious reason to suspect that untagged bugs would have markedly different
fix patterns than tagged ones.  However, false positives could well affect the relative frequencies of fault categories.

The \emph{Analyser} can detect and eliminate only a fraction of all non-fix IF-CC patterns.  However, the non-fix IF-CC hunks which have been considered in our analysis, are mostly classified in LRF:Other or broadly in LRF.  As a result, the actual frequency may fluctuate with our calculated frequency for LRF.  In general, given the pattern of our results and the results of our manual analysis, we do not believe that the non-eliminated non-fix IF-CC patterns have substantially affected our conclusions.
%\vspace{-2pt}
\subsection{External validity}
We have restricted our data set by selecting software systems implemented in Java which may not be representative for other software projects that are developed in other programming languages.  Since different programming languages have different constructs, the frequencies of different IF-CC fix patterns frequency may vary for different programming languages. 

We have consciously selected  well maintained software artifacts, with wide usage, which may not reflect the bug fix patterns in 
less well maintained and widely used open source software systems.  

Our artifacts are all open source software.  Proprietary artifacts may, or may not, show difference in frequencies of IF-CC fix patterns with our counted frequencies.  This suggests to look at proprietary software for future work.  

% We have classified \verb+if+ conditional changes particularly examining the implications for  MC/DC testing.  MC/DC testing technique is a 
% technique particularly used in embedded software system.  We do not know whether the bug fix patterns for embedded software 
% will reflect those in our artifacts.  
%\vspace{-2pt}
\section{Related works}\label{Related:work}

Historical information and relative frequencies of fault types can help practitioners
to select suitable testing methods.  From the testing researchers point of view, such historic information and relative frequencies may allow the general recommendation of particular 
unit testing techniques, or serve as inspiration to devise newer and more effective testing methods.  A number of researchers  (~\cite{Kim:2007:PFC:1248820.1248881,Ostrand:2010:SFP:1831708.1831743}) have devised techniques using historical information to identify the most fault prone files or modules of a software project.  These types of techniques are helpful to reduce testing effort by predicting mostly fault prone files, however, they do not provide much information about how to test them.  An attempt has been taken by Hayes in~\cite{Hayes94testingof}, where a number of fault classification studies are used to analyse the merits of various testing techniques in object oriented software.  But no relative frequency has been considered and the classification was largely based on the author's personal testing experience.     

There are several attempts to devise effective fault classification scheme for a number of different purposes (such as improving debugging).  One well-known fault classification effort was by Knuth~\cite{Knuth:Error} who classified the errors found or reported in ten years of the development of his well-known typesetting software, \TeX\, into nine categories.  The errors were classified manually by Knuth based on his own logs and recollections.  \TeX\ is a somewhat unusual software system that has been developed by Knuth himself, largely alone, to satisfy a specification he himself devised.   This classification is neither convenient to replicate without manual developer involvement, nor provides sufficient information to guide white-box testing. 

%Eisentadt~\cite{Eisenstadt:Bug} classified  bugs collected from a survey among some professional developers of the ``worst bugs'' they had %encountered of their professional career.  This study provides the cause and effect of some of the worst bugs.  However, it is quite broad and does not provide sufficient detail to select specific testing techniques.  According to this classification, the most common cause of ``worst bugs'' were memory allocation which is less severe in higher level programming language (like- Java) than languages (like- C/C++) which require manual memory management.

Static checking tools, such as FindBugs~\cite{Hovemeyer:FindBugs}, automatically classify some bug patterns.  These patterns indicate  bugs which occur due to mistakes with code idioms, or misuse of language features.  Those bug fix patterns can be detectable by using static checking tools.  As testing researchers, we are primarily interested to look for bug fix patterns  which demand testing rather than static checking to detect them - static checkers should be used to find and remove such bugs as can be identified \emph{before} testing!

DeMillo and Mathur~\cite{DeMillo:Classification} devised an alternative classification scheme based on semantic changes of bug fixing for the purpose of strengthen debugging and choosing suitable testing but their study was limited to a single software artifact.  In the recent time, Pan \etal~\cite{Pan:BugPattern} devised an automatic syntactic classification scheme for the better understanding about bugs and their frequency characteristics, and justified the scheme over seven open source Java software projects (not the same as ours).  Nath \etal ~\cite{Robert:FaultClassAnalysis} replicated this classification scheme on a single additional open source Java software artifact and suggested some potential improvements to the classification.  They also described the necessity of exploring more specific information for specific bug fix patterns to determine \emph{how} best to test for revealing the most common patterns.  
%
%Collating our previous study and Pan \etal's study, we found the single largest bug fix category is IF-CC (\verb+if+ condition change) patterns.   Pan \etal\  also classified the IF-CC fix patterns into six categories but they concluded that IF-CC fix patterns resulted in a trend of ``increased complexity of conditionals''.

In this study, we have devised an orthogonal fault classification scheme and calculated their frequencies over seven open source Java projects.  We have considered implications of this result in the context of testing \verb+if+ conditionals.

%\vspace{-2pt}
\section{Conclusion and Future Works}\label{Conclusion:Future}
The contributions of our paper are 14 automatically extractable \verb+if+ conditional bug fix patterns and their relative frequencies over seven open source Java software artifacts. We calculated the statistical significance by computing rank correlation and found that the frequencies of 10 major IF-CC fix patterns are highly correlated over seven software projects.  The classification is orthogonal and shares properties of ``pre-defined'' fault classes of existing specification based testing.  Considering the difficulties of generating specification based test sets for \verb+if+ conditionals, we have subcategorised the LRF into five sub-patterns.  However, we did not find significant cross-project correlation among the LRF sub-patterns.  We also have identified a set of non-fix patterns in IF-CC fix patterns.  We have accounted the fault frequency information for better understanding of testing \verb+if+ conditionals.

There are some obvious opportunities for extension of this work.  Most straightforwardly,
similar methodologies could be applied to other conditional constructs such as \verb+for+, \verb+while+, and \verb+do-while+ loops.  It is unknown, but certainly a relevant question, to see whether
the fault category frequency is similar in loop conditions and if-conditions.  There are specific patterns which might reasonably be suspected to be more common in loop conditions, such as changes to operators or a single atom to fix fencepost errors, so empirical data to support this long-held 
supposition would be of considerable practical interest.

The relatively common Multiple Fault (MF) pattern detection is also interesting for further study.
Some research has examined (e.g. ~\cite{Lau06:p-compsac-403}, ~\cite{Lau07:rst55}, ~\cite{Lau07:qsic117}) specification-based testing for double
faults in Boolean specifications.  However, the present study does not provide a clear indication of the proportion of multiple faults - which may involve more than two faults - can be detected.  Further research is required
to determine whether there are such appropriate testing techniques for the most common multiple fault types.

Finally, many other procedural and object-oriented programming languages offer similar conditional
constructs, but they may differ in their detailed syntax and implementation.  It would be interesting to see whether
there is any difference in fault category frequencies in projects implemented in different programming languages.

If fault based testing is to be effective, we believe it must be based on a solid empirical 
understanding of the faults that are actually present in real-world software systems.  This paper represents
a step in that direction.
%This work offers the possiblility for us to explore the following research questions, as future works:  What more context specific information can be found in Multiple Faults (MF), from a specification based testing perspective? Are there any obvious bug fix patterns with other language constructs such as for, while, and do-while, relating to Boolean expressions? If so, what is their relative frequency over different projects?  Finally, are the relative frequencies of IF-CC fault classes, similar for projects written in other programming languages (such as C/C++)? 

% After investigating the above questions, we intend to do following works: a) Devise a comprehensive fix pattern scheme for choosing suitable fault-based testing for each pattern, b) Justify the effectiveness of different testing techniques, considering the fix pattern frequencies, and c) Choose (or devise) most suitable bug fix identification technique, to get the noise free (at best) bug fix hunks.   
%\begin{itemize}
%\item We intend to analyze the multiple conditions of IF-CC fix patterns to provide more context specific information to the tester.
%\item We want to analyse the fix hunks related to the change of other conditions inside \emph{while, Switch, do-while, conditional} and \emph{for} rather than \verb+if+, in the perspective of boolean expression. 
%\item We further want to apply the IF-CC fix pattern classification procedure into the other programming language to see how the IF-CC fault classes vary on other programming language constructs. 
%\end{itemize}
%\small
\bibliographystyle{IEEEtran}
\let\oldbibliography\thebibliography
\renewcommand{\thebibliography}[1]{%
  \oldbibliography{#1}%
  \setlength{\itemsep}{4pt}%
}
\bibliography{if-patterns} 

\begin{thebibliography}{10}
\providecommand{\url}[1]{#1}
\csname url@samestyle\endcsname
\providecommand{\newblock}{\relax}
\providecommand{\bibinfo}[2]{#2}
\providecommand{\BIBentrySTDinterwordspacing}{\spaceskip=0pt\relax}
\providecommand{\BIBentryALTinterwordstretchfactor}{4}
\providecommand{\BIBentryALTinterwordspacing}{\spaceskip=\fontdimen2\font plus
\BIBentryALTinterwordstretchfactor\fontdimen3\font minus
  \fontdimen4\font\relax}
\providecommand{\BIBforeignlanguage}[2]{{%
\expandafter\ifx\csname l@#1\endcsname\relax
\typeout{** WARNING: IEEEtran.bst: No hyphenation pattern has been}%
\typeout{** loaded for the language `#1'. Using the pattern for}%
\typeout{** the default language instead.}%
\else
\language=\csname l@#1\endcsname
\fi
#2}}
\providecommand{\BIBdecl}{\relax}
\BIBdecl

\bibitem{Larry:Fault}
L.~J. Morell, ``A theory of fault-based testing,'' \emph{IEEE Transactions on
  Software Engineering}, vol.~16, pp. 844--857, August 1990.

\bibitem{Pan:BugPattern}
K.~Pan, S.~Kim, and {E. J. Whitehead Jr.}, ``Toward an understanding of bug fix
  patterns,'' \emph{Empirical Software Engineering}, vol.~14, no.~3, pp.
  286--315, August 2008.

\bibitem{Chilenski:MCDC}
J.~J. Chilenski and S.~P. Miller, ``Applicability of modified
  condition/decision coverage to software testing,'' \emph{Software Engineering
  Journal}, vol.~9, no.~5, pp. 193 --200, September 1994.

\bibitem{YuLauChen:j-jss-06-1}
Y.~T. Yu, M.~F. Lau, and T.~Y. Chen, ``Automatic generation of test cases from
  {B}oolean specifications using the {MUMCUT} strategy,'' \emph{Journal of
  Systems and Software}, vol.~79, no.~6, pp. 820--840, June 2006.

\bibitem{LauYu:j-tosem-05}
M.~F. Lau and Y.~T. Yu, ``An extended fault class hierarchy for
  specification-based testing,'' \emph{ACM Transactions on Software Engineering
  and Methodology}, vol.~14, no.~3, pp. 247--276, 2005.

\bibitem{ClassJung}
Y.~Jung, H.~Oh, and K.~Yi, ``Identifying static analysis techniques for finding
  non-fix hunks in fix revisions,'' in \emph{DSMM '09: Proceeding of the ACM
  first international workshop on Data-intensive software management and
  mining}.\hskip 1em plus 0.5em minus 0.4em\relax New York, NY, USA: ACM, 2009,
  pp. 13--18.

\bibitem{Class8}
A.~Mockus and L.~G. Votta, ``Identifying reasons for software changes using
  historic databases,'' in \emph{ICSM '00: Proceedings of the International
  Conference on Software Maintenance (ICSM'00)}.\hskip 1em plus 0.5em minus
  0.4em\relax Washington, DC, USA: IEEE Computer Society, 2000, p. 120.

\bibitem{Class9}
M.~Fischer, M.~Pinzger, and H.~Gall, ``Populating a release history database
  from version control and bug tracking systems,'' in \emph{ICSM '03:
  Proceedings of the International Conference on Software Maintenance}.\hskip
  1em plus 0.5em minus 0.4em\relax Washington, DC, USA: IEEE Computer Society,
  2003, p.~23.

\bibitem{Class4}
K.~Ayari, P.~Meshkinfam, G.~Antoniol, and M.~Di~Penta, ``Threats on building
  models from {CVS} and bugzilla repositories: the mozilla case study,'' in
  \emph{CASCON '07: Proceedings of the 2007 conference of the center for
  advanced studies on Collaborative research}.\hskip 1em plus 0.5em minus
  0.4em\relax New York, NY, USA: ACM, 2007, pp. 215--228.

\bibitem{Unidiff:web}
{The Open Group}, ``Diff description, the open group base specifications issue
  7,'' http://pubs.opengroup.org/onlinepubs/9699919799/utilities/\\diff.html,
  access date: 2011-03-17.

\bibitem{10.1109/TSE.2007.70731}
B.~Fluri, M.~Wuersch, M.~PInzger, and H.~Gall, ``Change distilling:tree
  differencing for fine-grained source code change extraction,'' \emph{IEEE
  Transactions on Software Engineering}, vol.~33, pp. 725--743, 2007.

\bibitem{JDT:web}
``Eclipse {J}ava {D}evelopment {T}ools ({JDT}) (project page),''
  http://http://www.eclipse.org/jdt/, access date: 2011-03-21.

\bibitem{Spearman:Wiki}
``Spearman's rank correlation coefficient,''
  http://en.wikipedia.org/wiki/Spearman's\_rank\_correlation\_coefficient,
  access date: 2011-02-04.

\bibitem{holm-bonferroni}
S.~Holm, ``A simple sequentially rejective multiple test procedure,''
  \emph{Scandinavian Journal of Statistics}, vol.~6, pp. 65--70, 1979.

\bibitem{Kim:2007:PFC:1248820.1248881}
\BIBentryALTinterwordspacing
S.~Kim, T.~Zimmermann, E.~J. Whitehead~Jr., and A.~Zeller, ``Predicting faults
  from cached history,'' in \emph{Proceedings of the 29th international
  conference on Software Engineering}, ser. ICSE '07.\hskip 1em plus 0.5em
  minus 0.4em\relax Washington, DC, USA: IEEE Computer Society, 2007, pp.
  489--498. [Online]. Available: \url{http://dx.doi.org/10.1109/ICSE.2007.66}
\BIBentrySTDinterwordspacing

\bibitem{Ostrand:2010:SFP:1831708.1831743}
\BIBentryALTinterwordspacing
T.~J. Ostrand and E.~J. Weyuker, ``Software fault prediction tool,'' in
  \emph{Proceedings of the 19th international symposium on Software testing and
  analysis}, ser. ISSTA '10.\hskip 1em plus 0.5em minus 0.4em\relax New York,
  NY, USA: ACM, 2010, pp. 275--278. [Online]. Available:
  \url{http://doi.acm.org/10.1145/1831708.1831743}
\BIBentrySTDinterwordspacing

\bibitem{Hayes94testingof}
J.~H. Hayes, ``Testing of object-oriented programming systems {(OOPS)}:a
  fault-based approach,'' in \emph{Object-Oriented Methodologies and Systems,
  volume LNCS 858}.\hskip 1em plus 0.5em minus 0.4em\relax Springer-Verlag,
  1994, pp. 205--220.

\bibitem{Knuth:Error}
D.~E. Knuth, ``The errors of {TeX},'' \emph{Software-Practice and Experience},
  vol.~19, no.~7, pp. 607--685, July 1989.

\bibitem{Hovemeyer:FindBugs}
D.~Hovemeyer and W.~Pugh, ``Finding bugs is easy,'' \emph{ACM SIGPLAN Notices},
  vol.~39, no.~12, pp. 92--106, December 2004.

\bibitem{DeMillo:Classification}
R.~A. Demillo and A.~P. Mathur, ``A grammar based fault classification scheme
  and its application to the classification of the errors of {TeX},'' Software
  Engineering Research Center; and Department of Computer Sciences; Purdue
  University, Tech. Rep., 1995.

\bibitem{Robert:FaultClassAnalysis}
\BIBentryALTinterwordspacing
S.~K. Nath, R.~Merkel, and M.~F. Lau, ``An analysis of fault classification
  scheme for {Java} software,'' Center for Software Analysis and Testing; and
  Faculty of ICT; Swinburne University of technology, Tech. Rep., May 2010.
  [Online]. Available:
  \url{http://www.swinburne.edu.au/ict/research/sat/technicalReports/TC2010-002.pdf}
\BIBentrySTDinterwordspacing

\bibitem{Lau06:p-compsac-403}
M.~F.~Lau, Y.~Liu, and Y.~T.~Yu, ``On detection conditions of double faults
  related to terms in {B}oolean expressions,'' in \emph{Proceedings of COMPSAC
  2006: the 30th Annual International Computer Software and Application
  Conference}, Sep. 2006, pp. 403--410.

\bibitem{Lau07:rst55}
------, ``On the detection conditions of double faults related to literals in
  {B}oolean expressions,'' in \emph{Proceedings of 12th International
  Conference on Reliable Software Technologies - Ada-Europe 2007}, ser. LNCS,
  no. 4498, Ada-Europe.\hskip 1em plus 0.5em minus 0.4em\relax ,, June 2007,
  pp. 55--68.

\bibitem{Lau07:qsic117}
------, ``Detecting double faults on term and literal in {B}oolean
  expressions,'' in \emph{Proceedings of 7th International Conference on
  Quality Software (QSIC~2007)}, Oct. 2007, pp. 117--126.

\end{thebibliography}
\end{document}